\documentstyle[prl,aps,epsfig]{revtex}
\begin{document}
\title{Conformations of Proteins in Equilibrium}
\author{Cristian Micheletti$^1$, Jayanth R. Banavar$^2$
and Amos Maritan$^1$}
\vskip 0.3cm
\address{(1) International School for Advanced Studies (S.I.S.S.A.)
and INFM, Via Beirut 2-4, 34014 Trieste, Italy}
\address{(2) Department of Physics and Center for Materials Physics,
104 Davey Laboratory, The Pennsylvania State University, University
Park, Pennsylvania 16802}
\date{\today}
\maketitle

\vskip 0.5cm

\begin{abstract}
We introduce a simple theoretical approach for an equilibrium study of
proteins with known native state structures. We test our approach
with results on well-studied globular proteins, Chymotrypsin Inhibitor
(2ci2), Barnase and the alpha spectrin SH3 domain and present evidence
for a hierarchical onset of order on lowering the temperature with
significant organization at the local level even at high temperatures.
A further application to the folding process of HIV-1 protease shows
that the model can be reliably used to identify key folding sites
that are responsible for the development of drug resistance .
\end{abstract}


Recent experimental and theoretical advances \cite{Baker_nat} have
shown that the topology of the native structure of a protein plays an
important role in determining many of its attributes. The number of
distinct native state conformations of proteins is limited
\cite{Chothia} -- often several distinct sequences fold into the same
       native state structure. The native state structures of proteins
contain secondary motifs (helices and sheets) in lower dimensional
manifolds which are curled into neat patterns (somewhat analogous to
the packing of clothes in a suitcase) and play a central role in the
folding process \cite{M2,Maritan1,Maritan2}.

The problem of protein folding entails the study of the
non-equilibrium dynamics in a rugged free energy landscape
\cite{funnel}. A valuable starting point for attacking such a problem
is through a thorough equilibrium analysis of proteins with known
native state structures. This would be useful for the determination of
the folding transition temperature by, for example, monitoring the
temperature at which thermodynamic quantities such as the specific
heat show a peak.  Such a study would lead to clear indications of the
equilibrium conformations of proteins as a function of the temperature
and provide a detailed picture of the onset of native-state like order
on lowering the temperature through the folding transition
temperature. Ideally, one would like information on the non-native
contacts and their role in facilitating the onset of native-state
ordering. Furthermore, it would be useful to incorporate amino-acid
specific interactions whenever such information is available.

At the present time, there is no simple theoretical framework for
accomplishing all these objectives. Go-like models \cite{Go2} have
proved to be useful for incorporating the role of the topology of the
native state structure in the folding process. In such models, one
ascribes a favorable attractive energy to the native contacts. There
have been numerous studies of Go-like models, which capture the notion
of minimal frustration \cite{funnel}, but even here, for realistic
off-lattice calculations, it is hard, if not impossible, to carry out
a full exploration of phase space in order to deduce equilibrium
averages. As mentioned previously, the ruggedness of the free energy
landscape carves out only a small part of phase space that the system
tends to be in leading to non-ergodic behavior even for modest size
proteins. Commonly used dynamics such as Monte-Carlo or molecular
dynamics tend to result in the system being compartmentalized in phase
space because of barriers that are difficult to surmount as the
temperature is lowered.

Recent progress has been made in the development of physically
motivated topology-based models
\cite{M2,eaton,Finkel,baker,Clem,Flam}.  The models vary greatly in
complexity and analytic tractability.  The only energy
contributions are postulated to arise from the establishment of native
interactions and therefore it is impossible to recover information on
non-native
contacts. In addition, a huge reduction in phase space is achieved
by introducing suitable constraints on contiguous  spin
variables along the chain. As a consequence, the ``true'' hiearchical
formation of the native state structure may lead to incompatibilities
with the phase-space constraints.

In this paper, we present a simple model for calculating the
equilibrium properties of heteropolymers or proteins with known native
states. The model builds on the importance of the native state
topology by assigning an attractive interaction between nearby amino
acids that are known to make native state contacts. It is also
possible to incorporate amino acid specific interactions into the
model. While the model, in its present simple form, does not
accurately represent the effects of self-avoidance \cite{ftn_sa}, it
nevertheless ensures that the native state is the true ground state
and satisfies all the steric constraints. The connectedness of the
chain as well as its entropy are captured in a simple, but
non-trivial, manner. The most significant advantage of the model is
that it can be used to explore the equilibrium thermodynamics without
being hampered by inaccurate or sluggish dynamics. A self-consistent
approximation is used to reduce the model to a Gaussian \cite{Dill}
form. The latter lends itself to a straightforward determination of
equilibrium quantities that identify the key folding sites \cite{M2},
such as those targeted by drugs against viral enzymes
\cite{BIOCH88,Cecco}.

The conformation of a protein is specified by the location of the
$C^{\alpha}$ atom of the $i$-th amino acid in sequence, $\vec{r}_{i}$.
In the native state, $\vec{r}_{i} \equiv \vec{r}^0_{i}$. A simple effective
Hamiltonian (energy function) that captures the essential features of
proteins is:

\begin{equation}
\tilde{{\cal H}} = {T \over 2} \,
K \sum_{i=1}^{N-1} ( \vec{r}_{i,i+1} - \vec{r}^0_{i,i+1} )^2
+ \sum_{i,j} { \Delta_{i,j} \over 2}\, X_{i,j} \,
\theta \left[-X_{i,j} \right] \ ,
\label{eq:ham}
\end{equation}

\noindent where $\vec{r}_{i,j} \equiv \vec{r}_i -\vec{r}_j$, $T$ is
the temperature ($K_B=1$), $\theta$ is the step function,

\begin{equation}
\theta(x) = \left\{
\begin{array}{l   r}
1 & \mbox{if $x > 0$\ ,}\\
0 & \mbox{otherwise\ ,}\\
\end{array} \right.
\end{equation}

\noindent $\Delta$ is the  contact matrix, whose element $\Delta_{ij}$
is 1 if residues $i$ and $j$ are in contact in the native state
(i.e. their C$_\alpha$ separation is below the cutoff $c=6.5$ \AA)
\cite{contact} and
0 otherwise, and

\begin{equation}
X_{i,j} = ( \vec{r}_{i,j} -\vec{r}^0_{i,j} )^2 -R^2  .
\end{equation}

The first term in the Hamiltonian involves harmonic interactions
between successive ``beads'' in the chain \cite{Doi}.  The temperature
factor ensures that the free-energy contributions of the peptide chain
are constant over the range of temperatures relevant to the folding
process. The second term in (\ref{eq:ham}) provides an attractive
interaction when amino acids which are in contact in the native state
conformation are in the proximity of each other
\cite{Go2}. Furthermore (\ref{eq:ham}) guarantees that a specific
native target structure is the ground state among all possible three
dimensional structures.  It is straightforward to generalize the
Hamiltonian so that all the pairwise interactions do not have the same
strength but are different and reflect the amino acid-specific
interactions.  In standard off-lattice approaches, the interaction
between non-bonded amino acids at a distance $d$, is taken to be a
square well potential, or some type of Lennard-Jones interaction.  Our
choice in Eq. (\ref{eq:ham}) is a sort of ``harmonic well'' which,
while being physically sound and viable, is suitable for a
self-consistent treatment, as explained below. The location of the
outer rim of the well is controlled by $R$, which can be set to a few
Angstroms ($R = 3$ \AA\ in the present study) to reflect the fact
that, when the separation of two residues is appreciably different
from the native one, their interaction is negligible. In its present
form, the model is complex and not amenable to a simple attack.  While
the first term has a simple quadratic form, the second term is
difficult to deal with because of the step function.

The key observation is that a dramatic simplification is accomplished
on making a self-consistent Gaussian approximation within which the
partition function is ${\cal Z} = \int \Pi_i d^3 r_i \ e^{-\beta {\cal
H}}$, with
\begin{equation}
{\cal H} = {T \over 2} K \sum_{i=1}^{N-1}
( \vec{r}_{i,i+1} - \vec{r}^0_{i,i+1} )^2
+ \sum_{i,j} {\Delta_{i,j} \over 2}\, X_{i,j}
\, p_{i,j} \ ,
\label{eq:hamapprox}
\end{equation}
\noindent where the $p_{i,j}$'s are determined self-consistently, in a
spirit similar to a local mean field approximation:
\begin{equation}
p_{i,j} \equiv
\langle \theta \left( R^2 -  ( \vec{r}_{i,j}
-\vec{r}^0_{i,j})^2 \right)\rangle_{\cal H} \ .
\label{eq:pij}
\end{equation}

Physically, $p_{i,j}$ represents the equilibrium probability of the
formation of the $i$-$j$ contact at temperature $T$.  $p_{i,{i+1}}$
can be conveniently frozen to 1 to reflect the strength of the peptide
chain. With the functional form given in (\ref{eq:hamapprox}), the
partition function can be readily deduced because all integrals are
Gaussian. The free energy, $F = - T \ \ln {\cal Z}$ is:

\begin{equation}
F = - {3N \over 2} \ln (2 \pi) - {3 \over 2} \ \ln
(\det {M}) -  {R^2 \over 2 T} \sum_{lm} \Delta_{lm} \ p_{lm}
\end{equation}

\noindent where the inverse matrix $M^{-1}$ is defined as

\begin{equation}
{{M}}^{-1}_{i,j} = \left \{
\begin{array}{r  l}
K (2 - \delta_{i,1} - \delta_{i,N}) + 2 \sum_l \Delta_{i,l}\, p_{i,l}/T \
\ &  {\rm for }\  i=j \\
-2 p_{i,j} \Delta_{i,j}/T + K \left[ - \delta_{i,j+1} -
\delta_{i,j-1}\right] \ \ & {\rm for }\ i \not= j \ .
\end{array}
\right.
\end{equation}

The self-consistency relations for the the probabilities $p_{ij}=
\langle \theta_{i,j} \rangle_{{\cal H}}$ are satisfied by finding the
fixed point of the recursion equation:

\begin{equation}
p^\prime_{ij} =  (2 \pi
G_{i,j}) ^{-3/2} \int d^3r \ e^{- r^2/ 2G_{i,j}} \ \theta(R^2 - r^2) .
\label{eq:rec}
\end{equation}

\noindent where the right-hand side depends on the $p_{ij}$ obtained
at the previous iteration through the matrix $M$, which enters the
definition of $G_{i,j} = {M}_{i,i}+ {M}_{j,j}- {M}_{i,j}-{M}_{j,i}$.
The solution of the recursion equations for the $p_{i,j}$,
(\ref{eq:rec}), entails the evaluation of incomplete $\Gamma$ functions
and converge to the fixed point very fast and typically an accuracy of
$10^{-4}$ is reached in a few dozen iterations.  Thus the Gaussian
nature of the Hamitonian allows a straightforward analytic attack on
the problem which, when combined with a rapidly convergent iterative
procedure on a computer, allows one to determine the equilibrium
properties of any protein with ease.

\begin{figure}
\begin{center}
\epsfig{figure=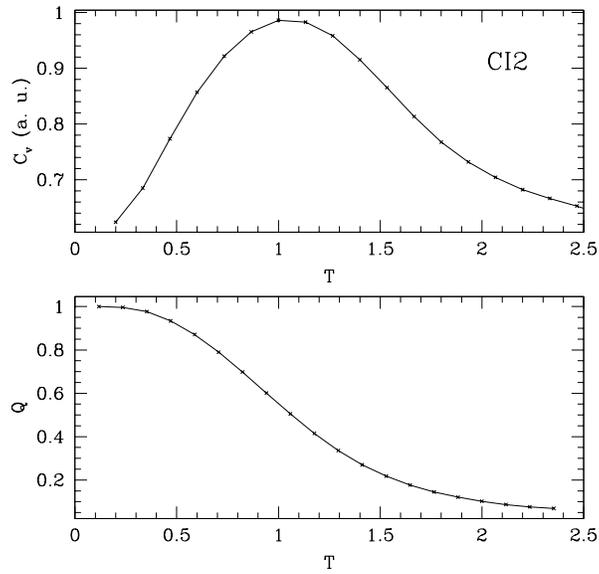,width=0.45\textwidth}
\end{center}
\caption{Plot of the specific heat (in arbitrary units) and the
native-state overlap as a function of temperature for the protein,
Chymotrypsin Inhibitor. The temperature is
measured in units of the folding transition temperature (identified
through the maximum of the specific heat).}
\label{fig:fig1}
\end{figure}

\noindent We present here the results for the globular proteins 2ci2,
Barnase and the $\alpha$-spectrin SH3 domain (PDB codes: 2ci2, 1a2p
and 1shg, respectively) for the simple case of uniform attractive
interactions between the amino acids which form the native contacts.
In all cases, it is straightforward to determine various thermodynamic
quantities as a function of temperature and identify \cite{scheraga}
the folding transition temperature as one at which the specific heat
exhibits a maximum (see e.g. Figure \ref{fig:fig1}). The width of the
specific heat peak at the folding transition in Figure \ref{fig:fig1}
is significantly smeared out compared to experiment \cite{Jackson} and
theory \cite{Kaya}. The cooperativity of the model can be easily
controlled by adjusting the value of $K$, with smaller values leading
to sharper ``transitions''. In addition, a change of $K$ also impacts
on the average amount of native structure that is formed at the native
state. Because we are particularly interested in characterizing the
progressive formation of native interactions, we chose the strength of
the peptide chain, $K$, by inspecting the behaviour of the fraction of
native contacts, $Q$, as a function of temperature.  $Q$, which is
often termed ``native-state overlap'', is defined as

\begin{equation}
Q = {\sum_{i,j}^\prime \Delta_{ij} \, p_{ij} \over
\sum_{i,j}^\prime \Delta_{ij}}
\end{equation}

\noindent where the prime denotes that the sum is not carried out over
consecutive pairs, in order to exclude the effects of the peptide
bond.  We find that, almost irrespective of the length of the target
proteins, a value of $K= 1/15$ yields $Q \approx 0.5$ at the folding
transition, consistent with experimental findings \cite{Pl} and
previous observations \cite{M2,Clem,Cecco,KarplusW,reaction}.

\begin{figure}
\begin{center}
\epsfig{figure=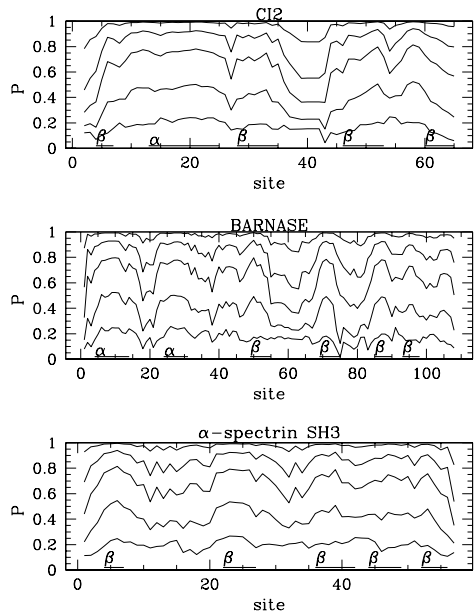,width=0.45\textwidth}
\end{center}
\caption{ Plot of $P_i$, the degree to which amino acid $i$ is in a
native-like conformation, versus $i$ for (a) 2ci2, (b) Barnase
and (c) $\alpha$-spectrin SH3. In
ascending order the curves are calculated at $T$= 2.0, 1.5, 1.0, 0.5
and 0.35 (measured in units of the folding transition temperature).
The bar at the bottom shows the secondary structure associated with
amino acid $i$.}
\label{fig:fig2}
\end{figure}

While $Q$ is a good global parameter to characterize the progress
towards the native state in a folding process, it is useful to monitor
the onset of native ordering at the level of individual residues. The
quantity, $P_i$,

\begin{equation}
P_i = {\sum_{j}^\prime \Delta_{ij} \, p_{ij} \over \sum_{j}^\prime
\Delta_{ij}}
\label{eqn:envir}
\end{equation}

\noindent provides an intuitive measure\cite{Finkel,karp} of the
degree to which amino acid $i$ is in its native-like
conformation. Figure \ref{fig:fig2} shows the profiles of such
environments for each amino acid in proteins CI2, barnase and the
$\alpha$-spectrin domain of SH3 as a function of the temperature.

In agreement with the experimental findings on these heavily
investigated proteins \cite{ci2b,barnase,sh3}, and also with
theoretical predictions\cite{Fersht,Rose1,cieplak}, we observe
that the secondary structures form at relatively high temperatures and
condition subsequent folding events.  Note the significant lack of
ordering of the loop regions even at the folding transition
temperature.  $\beta$-sheets are seen to form along one preferred
direction, while the formation of $\alpha$-helices occurs from the
ends. In general, the tendency of a site to reach its native
environment increases with both its degree of burial and the locality
of the contacts it forms. The intricate details of the figure reflect
an incremental assembly process and also the complex interplay between
the two effects mentioned above.

To further corroborate the validity of the proposed model in capturing
the important folding steps we consider an application to an important
enzyme, the protease of the HIV-1 virus (pdb code 1aid), which plays a
vital role in the spreading of the viral infection. Through extensive
clinical trials \cite{BIOCH88}, it has been established that there is
a well-defined set of sites in the enzyme that are crucial for
developing, through suitable mutations, resistance against drugs and
which play a crucial role in the folding process\cite{Cecco}.

\begin{figure}
\begin{center}
\epsfig{figure=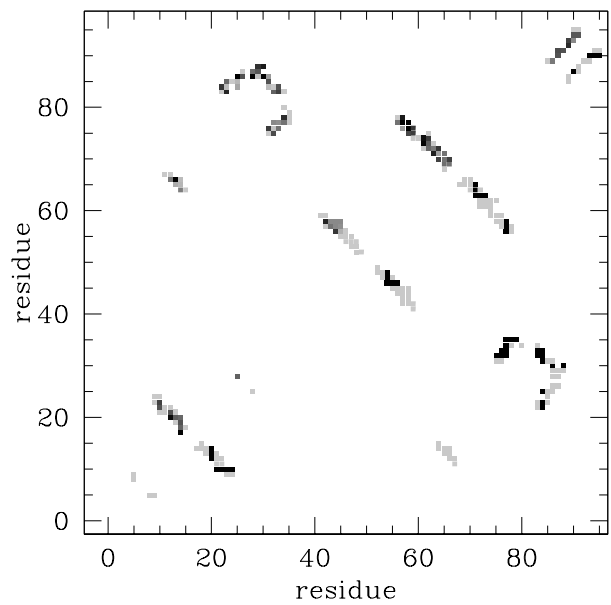,width=0.45\textwidth}
\end{center}
\caption{Contact map of HIV-1 PR monomer. Upper-left triangular
region: contacts with a large [small] formation temperature are shown
in dark [light] gray. Lower-right triangular region: contacts [not]
involving one or more of the known key mutating sites are shown in
dark [light] gray.}
\label{fig:C}
\end{figure}

To identify the key folding sites we looked for contacts that are
significantly formed above the folding transition temperature.
Quantitatively we define the formation temperature of a contact as one
at which a given $p_{i,j}$ takes on the value $0.5$. Our results are
summarized in Fig.~\ref{fig:C} (upper-left triangular region). As a
comparison, in the lower-right triangular region of the same figure,
we have highlighted the contacts involving the known mutating
sites. Remarkably, among the top 20 contacts with the largest
formation temperature there were 10 including one (or more) known
mutating site. A straightforward calculation shows that the
probability of observing this many successful matches by picking
contacts at random is less than 2 \%, confirming that our model
captures important aspects of the folding process with remarkable
precision and reliability.

In summary, the self-consistent Gaussian approach provides a simple
way of probing the equilibrium properties of proteins with the
incorporation of amino acid specific interactions (when known) shorn
away from the usual complications associated with imperfect or
inadequately studied dynamics.  A key advantage is that our approach
is essentially analytic and the quantities of interest may be determined
with arbitrary accuracy quite easily.

We are indebted to Alessandro Flammini for a careful reading of the manuscript
This work was supported by INFM, Murst Cofin99, and NASA.

\begin{thebibliography}{99}
\bibitem{ftn_sa} The model does not include a hard-core repulsion
between the amino acids. However, steric clashes are avoided to some
extent because of the incorporation of the geometry of the
self-avoiding native state.
\bibitem{contact} Alternatively, residues can be considered to be in contact
if they contain heavy atoms that are within 4-5 \AA. This alternative
choice does not significantly alter the results discussed here (data
not shown).
\end{thebibliography}
\end{document}